\documentclass[twocolumn,english,aps,prl,showpacs,superscriptaddress,
groupedaddress]{revtex4-1}
\usepackage[T1]{fontenc}
\usepackage[latin9]{inputenc}
\usepackage{graphics}
\usepackage{verbatim}
\usepackage{amsthm}
\usepackage{amsmath}
\usepackage{amssymb}
\usepackage[pdftex]{graphicx}
\usepackage{graphicx}
\usepackage{esint}
\usepackage{dcolumn}
\usepackage{bm}
\usepackage{ifthen}
\usepackage{babel}
\usepackage{color}
\usepackage{multirow}
\usepackage{ulem}

\usepackage[usenames,dvipsnames]{xcolor}

\usepackage{textcomp}

\makeatletter
\begin{document}
\title{Physics of Giant ElectroMagnetic Pulse generation in short pulse laser experiments}

\author{A. Poy\'e$^{1}$, S. Hulin$^{1}$, M. Bailly-Grandvaux$^{1}$, J.-L. Dubois$^{2}$,
J. Ribolzi$^{2}$, D. Raffestin$^{2}$, M. Bardon$^{2}$, F. Lubrano-Lavaderci$^{2}$, 
E. D'Humi\`eres$^{1}$, J. J. Santos$^{1}$, Ph. Nicola\"i$^{1}$, V. Tikhonchuk$^{1}$}
\affiliation{\vspace{2mm}
$^{1}$Centre Lasers Intenses et Applications, University Bordeaux, CNRS, CEA, Talence 33405, France}
\affiliation{$^{2}$CEA/DAM/CESTA, BP 12, Le Barp 33405, France}

\begin{abstract}
In this paper we describe the physical processes that lead to the generation of Giant Electro-Magnetic Pulses (GEMP) on powerful laser facilities. Our study is based on experimental measurements of both the charging of a solid target irradiated by an ultra-short, ultra-intense laser and the detection of the electromagnetic emission in the GHz domain. An unambiguous correlation between the neutralisation current in the target holder and the electromagnetic emission shows that the source of the GEMP is the remaining positive charge inside the target after the escape of fast electrons accelerated by the ultra-intense laser. A simple model for calculating this charge in the thick target case is presented. From this model and knowing the geometry of the target holder, it becomes possible to estimate the intensity and the dominant frequencies of the GEMP on any facility.
\end{abstract}
\maketitle

\section{Introduction}
The continuous progress made in the construction of high intensity laser opens new domains of physics and impacts applications from inertial confinement fusion to laboratory astrophysics and material processing. However the interaction of a high intensity laser pulse with a solid target is associated with the generation of intense broadband  electromagnetic pulses across a wide frequency range from radio frequencies \cite{Pearlman1977} to x-rays \cite{Courtois2009}. With the new generation of PetaWatt lasers, one would expect Giant Electro-Magnetic Pulses (GEMP) in the Giga- to the Tera-Hertz domain, exceeding 1~MV.m$^{-1}$ \cite{Mead2004,Brown2008} at distances of 1 m from the target, which could be very destructive for any electronic device. On the other hand, the process responsible for this violent emission, if properly controlled, can lead to the production of enormous quasi-static magnetic fields exceeding 1 kT \cite{Fujioka2013,Santos2014} in a 1~mm$^3$ volume, which presents exciting new opportunities for many applications or fundamental research fields such as particle guiding, atomic physics or magnetohydrodynamics.

In this paper, we present the physics of GEMP generation with short laser pulses and validate it unambiguously in dedicated experiments. In the experimental section, we will show that the laser pulse generates a localised charge $Q$ on the target by the escaped electrons. A discharge current $I$ comes from the ground through the holder to neutralize it. The GEMP is thereafter emitted by the target holder which behaves like an antenna where the discharge current oscillates. This explains two major aspects of the GEMP emission. The first is that the emission spectrum is defined by the geometry of the antenna composed of the target and its support. The second is that the strength of the GEMP emission is directly related to the amount of accumulated charge $Q$ in the target. Assuming the target-holder impedanceis known, the prediction of this charge  becomes crucial for experimental designs, either for GEMP mitigation or for new conceptions of strong field generation.\\
For this reason, in the second part of this article, we will present a simplified model to calculate the remaining positive charge $Q$ in the case of a thick target irradiated by an intense laser. We will discuss the domain in which our model is valid showing that it may apply to a broad range of intense laser facilities worldwide. \\
To conclude, we will discuss the opportunity of controlling the discharge process to either decrease or modulate the GEMP emission, or generate strong quasi-static magnetic fields.

\section{Target charging experiment}
\subsection{Experimental set-up}
\begin{figure}
\begin{centering}
\includegraphics[height=5.5cm]{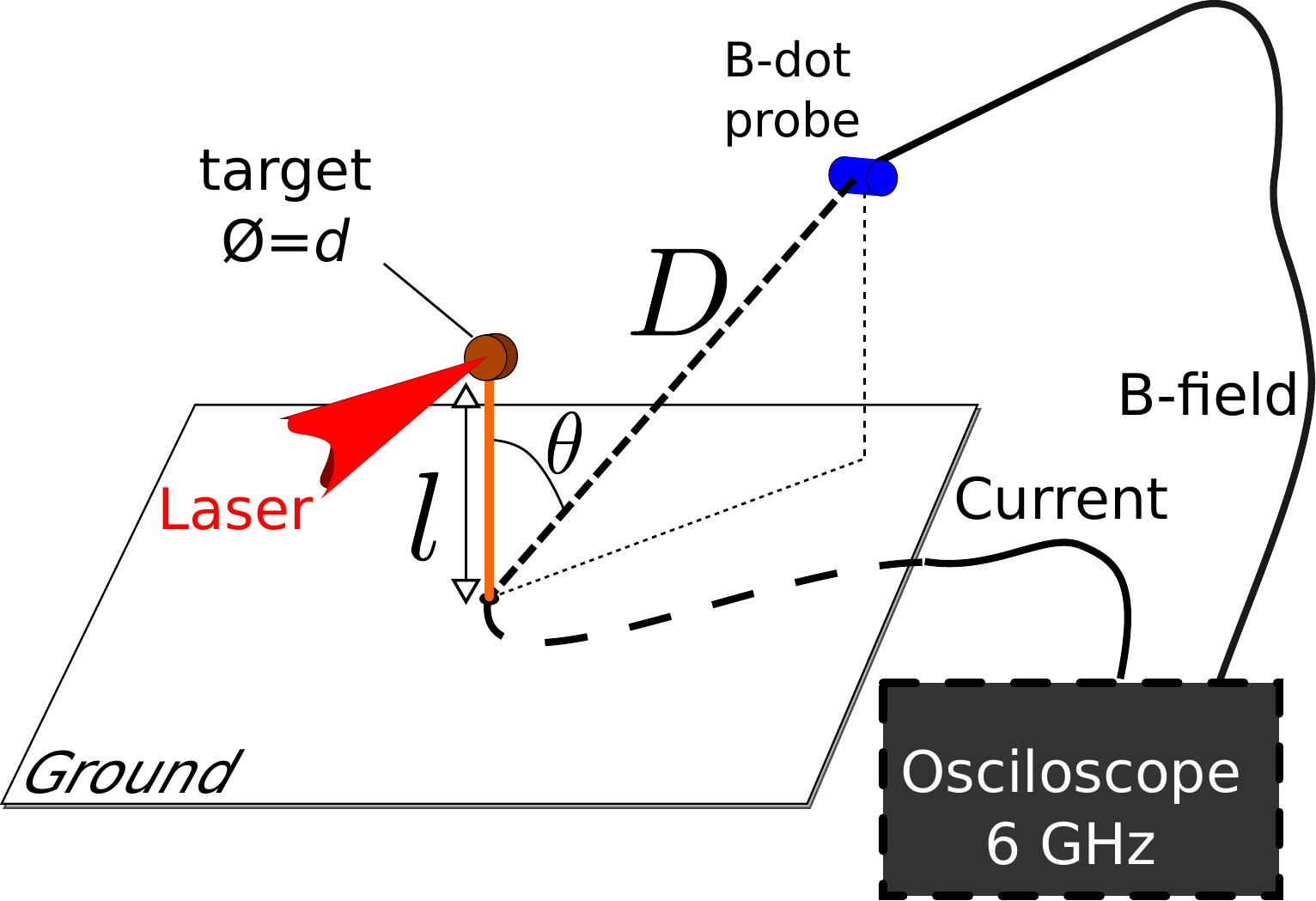}
\par\end{centering}
\caption{Experimental set-up: the target, its holder and the diagnostics.\label{fig:exp_scheme}}
\end{figure}

In our experiment, a Ti:Sapphire laser with a central wavelength of $807$~nm was focused down to a circular Gaussian focal spot of $12$~\textmu\-m FWHM (Full Width at Half Maximum) at normal incidence on the flat surface of a cylindrical ($d=10$~mm) thick ($3$~mm) target of either Copper, Aluminium or Tantalum. The laser temporal contrast was $10^{-7}$ and we varied both the energy on target [20~mJ$\rightarrow$80~mJ] and the pulse duration  [30~fs$\rightarrow$5~ps] for each type of target. As shown in Fig \ref{fig:exp_scheme}, the target was supported by a brass wire of $1$~mm diameter fixed to a connector located in the middle of a large metallic ground plane of $20$~cm~$\times$~20~cm. With such a system we could directly connect the brass wire to the core of a coaxial cable and fix the distance between the target side and the ground plate to $l=4.5$~cm. The coaxial cable was then connected with a $60$~dB attenuation to a $6$~GHz bandwidth oscilloscope to measure the recharge current $I$ circulating in the target. The integration of this current provides the charge of the target $Q_{exp}$ generated by the pulse. 

A magnetic probe "Bdot", set behind the target at a distance of $225$~mm  from the target center ($46$~mm top and $90$~mm right with respect to the laser axis), connected to the same $6$~GHz oscilloscope, measures the time derivative of the emitted magnetic field in the polarisation direction perpendicular to the brass wire axis. We record $B_{exp}$, the maximum of the emitted magnetic field obtained after integration.

\subsection{Experimental results}
\begin{figure}
\begin{centering}
\includegraphics[height=6.5cm]{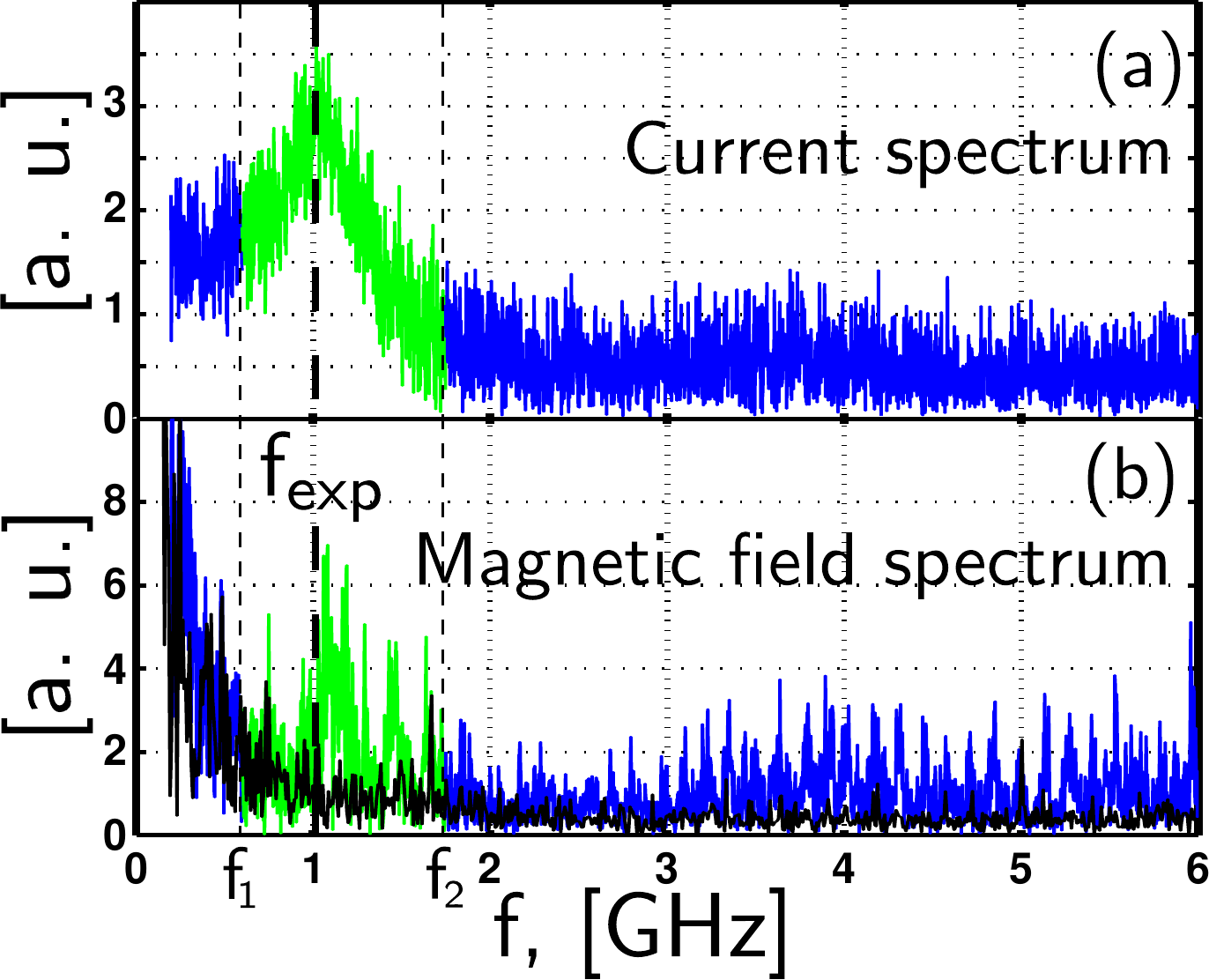}
\par\end{centering}
\caption{Typical signal spectra for a) the current and b) the magnetic field. $f_{exp}= 1.0\pm 0.1$ GHz, $f_1 = 0.6$~GHz and $f_2 = 1.7$~GHz for all experimental points. The black signal is the integration noise. \label{fig:spectra}}
\end{figure}

We start by focusing our attention on signal spectra. Figure \ref{fig:spectra}a is obtained with a FFT of the raw current signal and Figure \ref{fig:spectra}b is the FFT of the raw Bdot signal integrated in the Fourier space. Despite the various laser parameters and target materials used, all electric current and magnetic field spectra present the same main frequency $f_{exp} = 1.0 \pm 0.1$~GHz. This can be explained by considering the target-holder system as a dipole antenna where the ground plate plays the role of a mirror. In that case, the emission frequency is given by: $f_a\simeq c/4\left( l+\pi d/2\right)=1.2$~GHz, where the target-holder system is assimilated to a stalk of a length which equals the holder length plus half the target perimeter. This estimate corresponds fairly well to the measured frequency considering the antenna volume simplification. For a better definition of the emission spectrum, it is necessary to perform a numerical simulation such as those described in \cite{Dubois2014}, which reproduces the experimental signal.

\begin{figure}
\begin{centering}
\includegraphics[height=5.5cm]{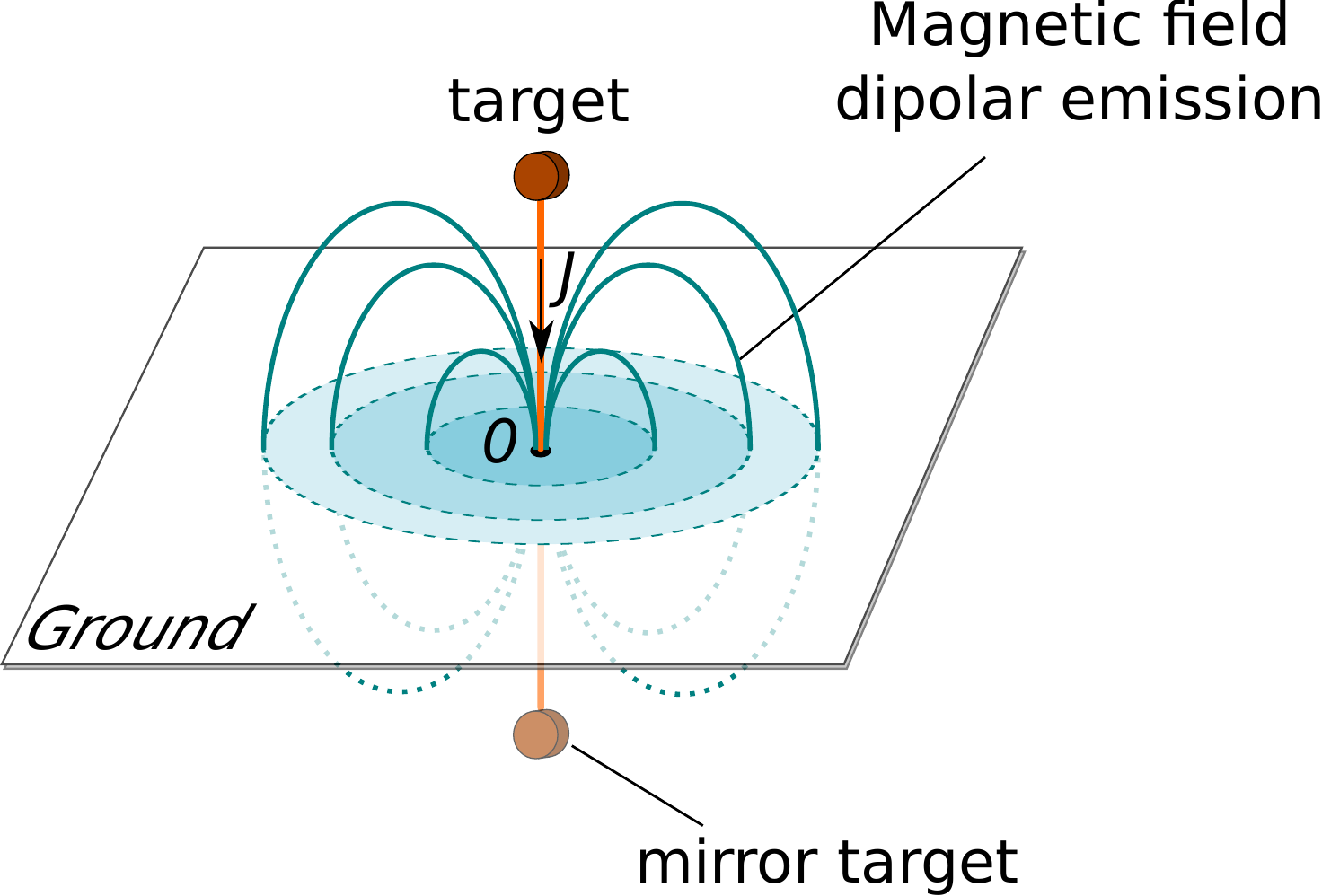}
\par\end{centering}
\caption{Scheme of the electro-magnetic emission from the holder. \label{fig:emission_scheme}}
\end{figure}

To have a better understanding of the electromagnetic emission, we also provide a movie of the electric field emitted during the hot electron ejection and the charge neutralisation, see Ref. \cite{movie}. The first spherical front is generated by the charge separation during electron ejection \cite{Felber2005, Sagisaka2008}. As the frequency range of this emission is in the THz range, it is out of the scope of our diagnostics and it is not critical for electronic devices. Note that the dipole formed by the positive charge on the target and the ejected electrons is oriented along the laser axis, as shown in \cite{movie}. The second emission front appears later, during the charge neutralisation. The dipole is now defined by the target-holder system and its reflection in the ground plate mirror. Figure \ref{fig:emission_scheme} presents a scheme of this emission, while the oscillations of the neutralisation current can be observed in the movie \cite{movie}.

\begin{figure}
\begin{centering}
\includegraphics[height=6.5cm]{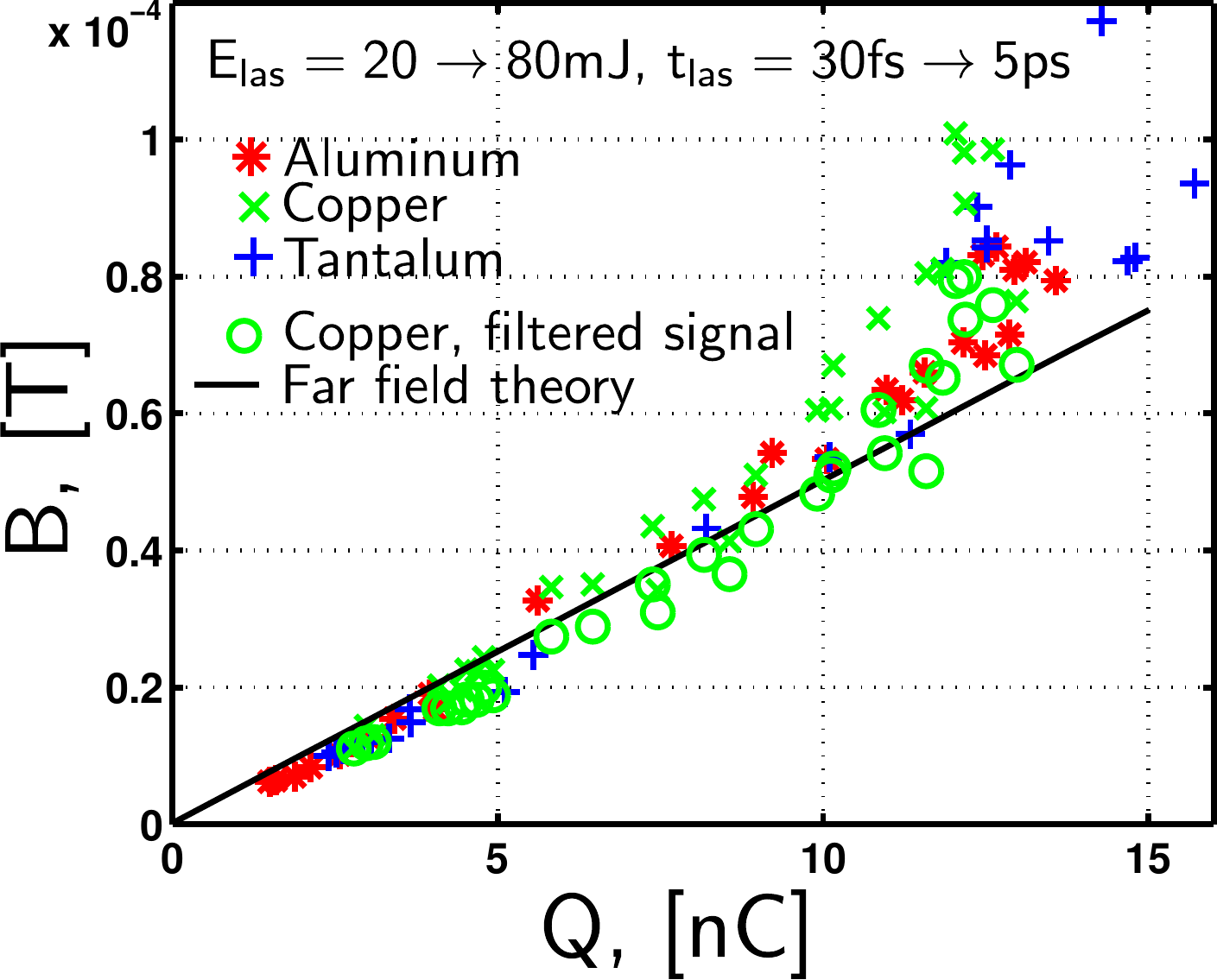}
\par\end{centering}
\caption{a) The maximum emitted magnetic field versus the associated target charge for several laser parameters and different target materials (copper, aliminum, tantalum), experiments and model (details in text). The experimental points are averaged over 5 shots. \label{fig:Q_vs_B}}
\end{figure}

We focus now on the amplitude of the electromagnetic field. The relation between $Q_{exp}$ and $B_{exp}$ is demonstrated in Fig. \ref{fig:Q_vs_B}a [\textcolor{blue}{+},\textcolor{red}{$\ast$},\textcolor{green}{$\times$}], for various laser parameters and target materials. As the spectra analysis and the movie suggest, GEMP are produced by a dipole antenna emission. The contact point between the ground plate and the brass wire is the center of the antenna and it is possible to estimate the amplitude of the magnetic field detected by the "Bdot" probe measurement using the far-field emission of a dipole antenna at a distance $D$ and at an angle $\theta$ with respect to the antenna axis \cite{Jackson1975}:
\begin{equation}
B_{a} =  \frac{\mu_0 I \cos \left(\frac{\pi}{2}\cos \theta \right) }{2\pi D \sin\theta}. \label{eq:B_th}
\end{equation}
In our case $D= 240$~mm, $\theta =77$\textdegree~and the discharge current is estimated by $I=1.6cQ_{exp}/\left( l+ \pi d/2 \right)$ where $1.6$ is the impedance adaptation coefficient. As the "Bdot" probe signal is noisy (in black in Fig. \ref{fig:spectra}b) and also contains cavity modes of the experimental chamber (in blue in Fig. \ref{fig:spectra}b), we filtered it between $f_1$ and $f_2$ to isolate the antenna component signal (in green in Fig. \ref{fig:spectra}b). The experimental filtered $B_{exp}$ ([\textcolor{green}{$\circ$}] in Fig. \ref{fig:Q_vs_B}) is compared to the magnetic field from Eq. (\ref{eq:B_th}) ([solid line] in Fig. \ref{fig:Q_vs_B}). The closed  agreement shows that the ground-holder-target system is the antenna, which emits GEMP and that the accumulated charge is the energy source of GEMP. Thus, a model was developped to estimate this accumulated charge.

\section{Target charging model}

\subsection{Model assumptions}
The evaluation of the charge $Q$ accumulated on the target after the end of a laser pulse is a multi-physical problem, which involves various effects related to laser-matter interaction, particle diffusion and collision in the target, electromagnetic properties and an overall description of the chamber environment \cite{Bateman2012, Dubois2014}. Here, we present a model, which estimates, to the nearest order of magnitude, the target charge assuming, that the laser absorption coefficient is known. It includes all major physical effects and is applied to thick metallic targets, i.e. the target thickness is larger than the range of accelerated electrons. Thus the electrons are ejected only from surface irradiated by the laser. The basic principles of the electron dynamics can be described as follows. The laser pulse accelerates electrons at the impacted target surface. They spread into the target dissipating their energy inside and creating an electric potential at the target surface \cite{Crow1975,Mora2003,Schreiber2006}. This potential is a barrier that the accelerated electrons must overcome in order to escape completely from the target. The number of escaped  electrons defines the net positive charge left on the target surface.

Instead of following the dynamics of the hot electrons, as was done in Ref. \cite{Dubois2014}, here we focus our attention on the maximum number of electrons that may escape. The basic assumptions concern the characteristics of the accelerated electrons. We assume that they are created in the laser focal spot of radius  $r_{\rm las}$, and their density equals approximately the relativistically corrected critical density $n_c=1.1\sqrt{1+{a_0}^2/2}\cdot 10^{21} \lambda_{\rm las}^{-2}$ cm$^{-3}$, with $\lambda_{\rm las}$ being the laser wavelength in $\mu$m, $a_0  =0.85\lambda_{\rm las}I_{\rm las}^{1/2}$ the dimensionless laser vector potential and $I_{\rm las}=0.65 E_{\rm las}/t_{\rm las}\pi r_{\rm las}^{2}$ the laser intensity with units of $10^{18}$~Wcm$^{-2}$ \cite{Gibbon2005}. The energy of the accelerated electrons follows a Maxwell-J\"uttner distribution function \eqref{eq:Max-Jutt_dist_funct} characterised by the hot electron temperature $T_h$:
\begin{equation}
f_e(\varepsilon_e)=\frac{1}{\mathcal{A}} \gamma_e p_e \exp (-\varepsilon_e/T_h), \label{eq:Max-Jutt_dist_funct}
\end{equation}
 where $p_e$ is the electron momentum, $\gamma_e$ the relativistic factor and $\mathcal{A}$ the normalisation factor. Such an approximation for the electron energy distribution is reasonable because the laser accelerated electrons are scattered by the electrons and ions in the target. For the temperature, we use Beg's empirical law \cite{Beg1997} in the interval $0.03\lesssim a_0\lesssim 1$ or ponderomotive scaling \cite{Wilks1992} for higher laser intensities:
\begin{equation}
T_h = m_e c^2 \max \bigg\{0.47\,a_0^{2/3}, \quad\sqrt{1+a_0^{2}}-1 \bigg\}. \label{eq:BegLaw}
\end{equation} 
For intensities below $10^{15}$~Wcm$^{-2}$ ($a_0 \lesssim 0.03$ ), the hot electron temperature is estimated from the model of laser collisional absorption \cite{Fabbro1985}, $T_h = 3m_e c^2 a_0^{4/3} $. The ratio between the absorbed laser pulse energy $\eta_{\rm abs} E_{\rm las}$ and the  mean energy of accelerated electrons $\langle \varepsilon_e\rangle=\int_{0}^{\infty}\varepsilon_e f_e\,d\varepsilon_e$ gives the total number of accelerated electrons:
\begin{equation}
N_e=\eta_{\rm abs} E_{\rm las}/\langle \varepsilon_e\rangle \label{eq:N-accelerated}
\end{equation}

\subsection{Potential barrier calculation}
The critical section of the model is the estimation of the potential barrier, which eventually defines the maximum number of escaped electrons. The first contribution to this barrier comes from the cloud of energetic electrons outside that target, which creates a charge separation potential $\phi_{th}$ as described in the ion acceleration model  \cite{Crow1975,Mora2003,Schreiber2006}. The high energy electrons may overcome the potential barrier and escape completely from the target. The remaining positive charge generates a global background potential $\phi_{bg}$, which spreads over the target surface.
First, we focus on the first part of the potential, $\phi_{th}$, which is described by the Poisson equation \eqref{eq:poisson} assuming a Boltzmann distribution of electrons in the potential barrier:
\begin{equation}
\epsilon_{0}\Delta\phi_{th}=-e\big(n_{i}-n_{c}\exp (e\phi_{th}/T_h)\big), \label{eq:poisson}
\end{equation}
where $e$ is the electron charge and $\epsilon_0$ is the vacuum dielectric permittivity. The ion density is described by a Heaviside function, $n_{i}=n_c H(x)$, where $x=0$ defines the target surface. This hypothesis is valid as long as the characteristic scale of the ion density is smaller than the hot electron Debye length defined below.

\begin{figure}
\begin{centering}
\includegraphics[height=9cm]{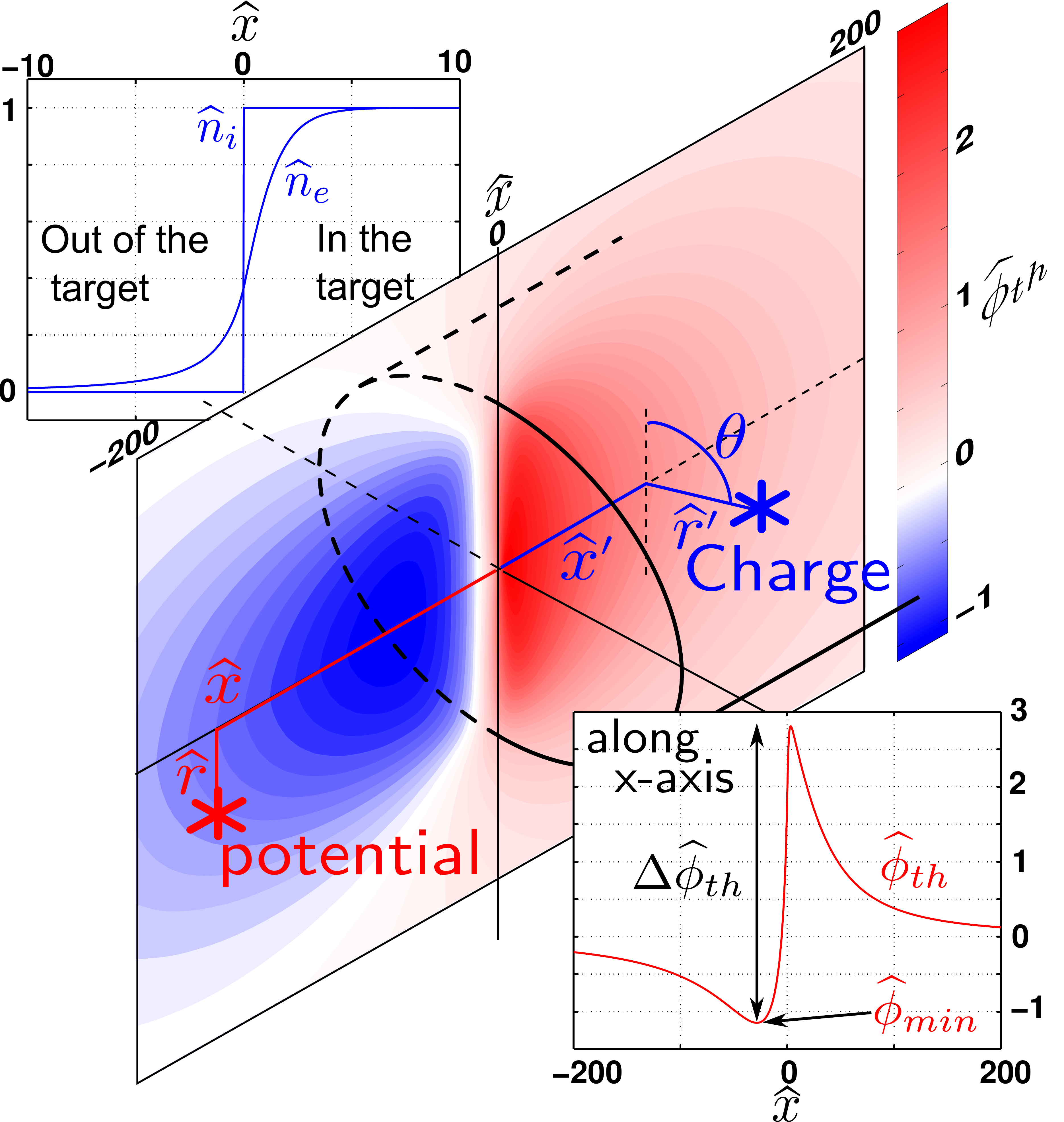}
\par\end{centering}
\caption{Example of a space charge reconstruction for the normalized radius $\widehat{R}=50$. Top inset: axial distribution of the ion and electron density from Eq. (\eqref{eq:ne}). Bottom inset: the axial potential.\label{fig:potential_built}}
\end{figure}

\begin{figure}
\begin{centering}
\includegraphics[height=6.5cm]{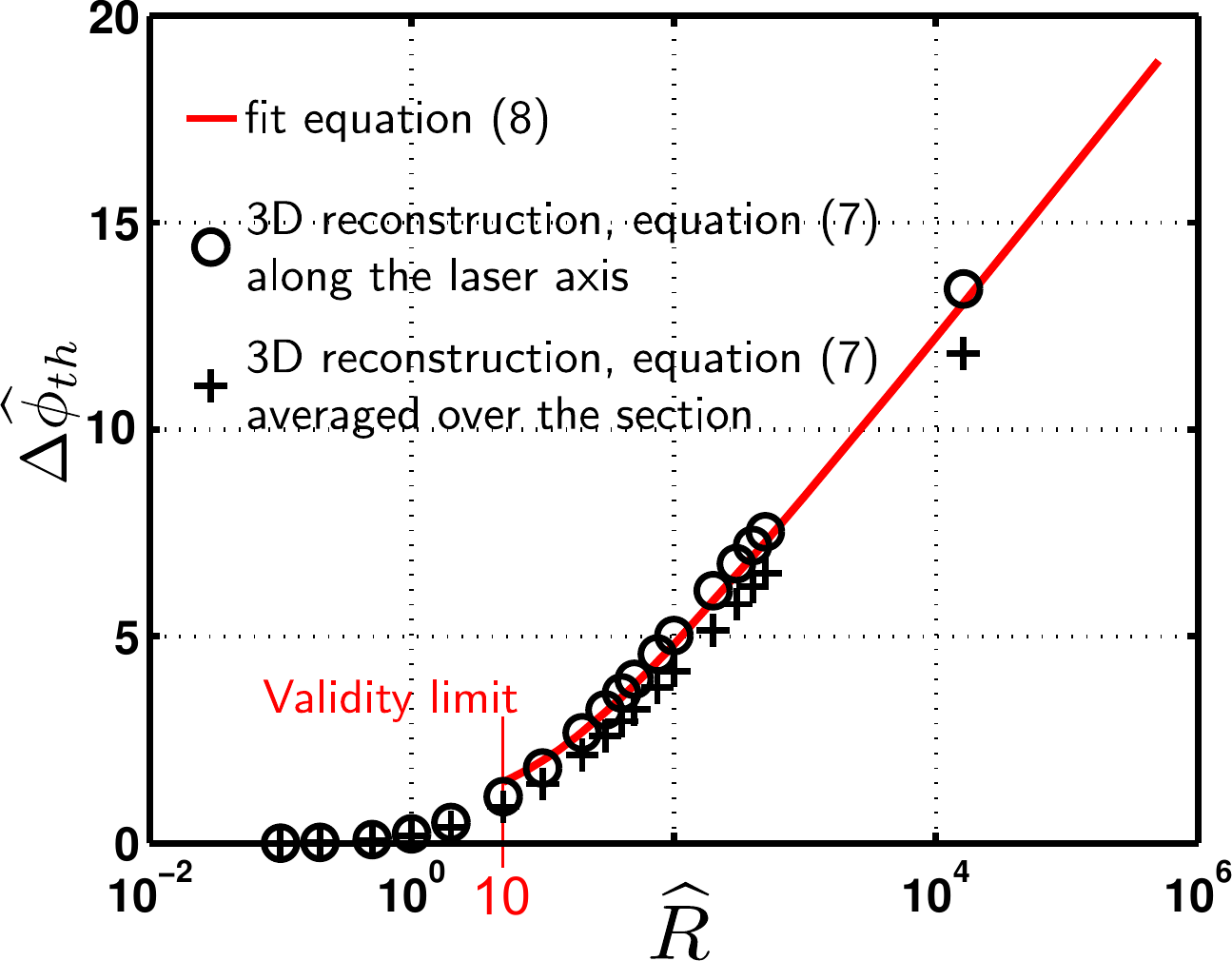} 
\par\end{centering}
\caption{Dependence of the maximum normalized potential barrier $\Delta\widehat{\phi}_{th}$ on the normalized laser spot radius $\widehat{R}$.\label{fig:potential_result}}
\end{figure}

This equation has a divergent solution for the potential in one dimension \cite{Carron1976}. Here, we present a method which determines completely the potential barrier: $\phi_{th}$ is convergent inside and outside the target. In contrast to the ion acceleration problem \cite{Crow1975,Mora2003,Schreiber2006}, where only the electric field at the surface matters, the problem of electron escape requires a knowledge of the whole potential profile. In the one-dimensional model, one obtains the following electron density distribution: 
\begin{equation}
n_{e}(x)=\left\{
\begin{array}{cll}
& n_c\,\exp\big(-\exp(-\xi x/\lambda_D)\big) \qquad x>0,\\
& n_c\left({\rm e}^{1/2}-x/\lambda_D\sqrt{2}\right)^{-2}  \qquad x<0,
\end{array}
\right. \label{eq:ne}
\end{equation}
where $\lambda_{D}=\sqrt{\epsilon_{0}T_h/n_c e^{2}}$ is the hot electron Debye length. The coefficient $\xi=0.9288$ is calculated to conserve the electro-neutrality. As the electron density \eqref{eq:ne} is a converging function: we can use it to determine the potential $\phi_{th}$ with a three-dimensional calculation. The hot electrons are distributed over a cylinder of radius $r_{\rm las}$ with the $x$-axis perpendicular to the target surface (see figure \ref{fig:potential_built}). The potential is numerically calculated from the charge distribution Eq. \eqref{eq:ne}:
\begin{equation}
\phi_{th}(r,x)= \frac{e}{4\pi\epsilon_0}\int \frac{ \big(n_{i}(x')-n_{e}(x')\big)\,dx'\,r'\,dr'\,d\theta}{\sqrt{(x-x')^{2}+r^{2}+{r'}^{2} -r' r\cos\theta}}.
\label{eq:phith_rebuild}
\end{equation}
It is important to note here that the potential normalized to the hot electron temperature, $\widehat{\phi}_{th}= e\phi_{th}/T_h$,  depends only on the normalized radius $\widehat{R}=r_{\rm las} /\lambda_D$. Once the potential $\widehat{\phi}_{th}\left(r,\theta,x\right)$ is calculated, we extract the normalized potential barrier labelled $\Delta\widehat{\phi}_{th}$. It is defined as the maximum potential variation over the x-axis, averaged over the cylinder section. For large values of $\widehat{R}>10$, we provide a fit for the potential barrier at the center of the target cylinder 
\begin{equation}
\Delta\widehat{\phi}_{th} \simeq 4\log_{R}\left(1 - \log_{R}/(1+\log_{R}^2 )\right) \label{eq:fit_phi}
\end{equation}
with $\log_{R}=\log_{10}\widehat{R}$. The potential barriers are presented in Fig. \ref{fig:potential_result}.

\subsection{Charge estimations}
We now combine the various aspects of the model to estimate the target charge after the laser pulse. Knowing the temperature \eqref{eq:BegLaw}, the normalized potential barrier \eqref{eq:fit_phi}, the distribution function \eqref{eq:Max-Jutt_dist_funct}, and the total number of hot electron \eqref{eq:N-accelerated}, one can readily determine the amount of charge escaping the potential barrier: 
\begin{equation}
Q =e N_{e}\int_{T_h\Delta\widehat{\phi}_{th}}^{\infty} f_e(\varepsilon_e)\,d\varepsilon_e. \label{eq:final_Q}
\end{equation}

The model's assumptions are valid for several laser facilities with various laser parameters, with the exception of high contrast systems where the electron acceleration may not be descripted by \eqref{eq:BegLaw}.

\subsection{Effect of the target size on the charge}
\begin{figure}
\begin{centering}
\includegraphics[height=6.5cm]{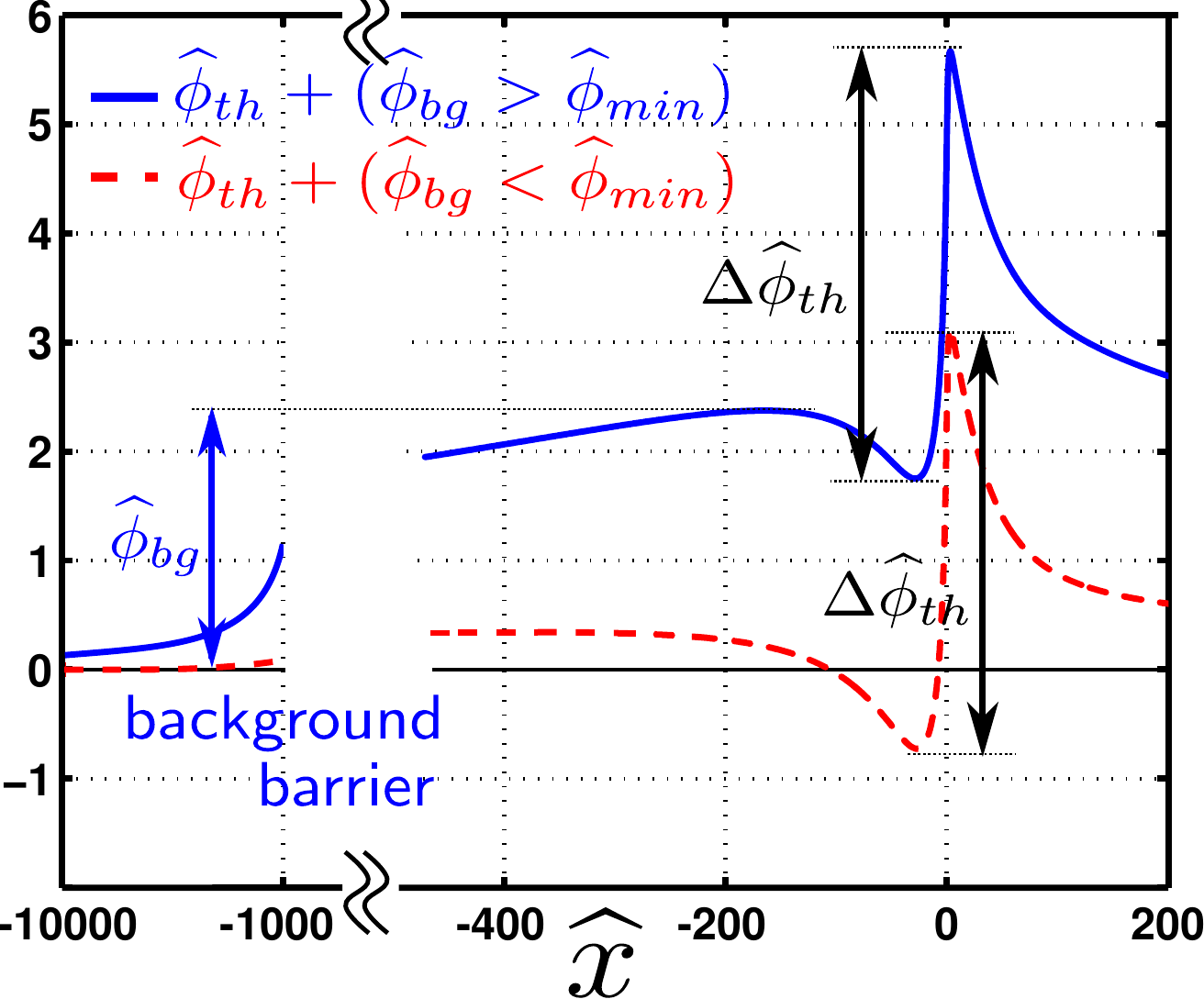}
\par\end{centering}
\caption{The potential for $\widehat{R}=50$ with an effective (solid blue line) or ineffective (dashed red line) arbitrary background potential $\widehat{d}=d/\lambda_D=1000$.\label{fig:potential_bg}}
\end{figure}

We focus now on the background potential $\phi_{bg}$, which is due to the uncompensated positive charge left at the target surface by the escaped electrons. This positive charge spreads over the whole target surface, which reveals a target size effect. For infinite or big enough targets, the background potential remains negligible because the surfacic charge is low. For small targets, the charge concentrates at the target surface and develops a background potential, which may overcome the potential barrier. To investigate the impact of the background potential on the target charge, we estimate $\phi_{bg}$ as follows:
 \begin{equation}
\phi_{bg} =\frac{Q}{2\pi\epsilon_0 d}. \label{eq:phi_bg}
\end{equation} 
  This formula assumes a uniformly charged disc of a negligible thickness compared to its radius. In the domain of practical interest, the disc radius is much larger than the hot electron localisation zone, which implies that $\phi_{bg}$ is fairly constant along the laser axis in the $\phi_{th}$ barrier range. As a consequence, the whole potential curve $\phi_{th}$ is offset by the background potential. Two examples of the total normalized potential $ \widehat{\phi}_{th}+\widehat{\phi}_{bg}$ are shown in Fig. \ref{fig:potential_bg}.
If $\widehat{\phi}_{bg}< \widehat{\phi}_{min}$, the potential barrier value is unchanged (red dashed line).  If $\widehat{\phi}_{bg}> \widehat{\phi}_{min}$, the potential barrier increases (blue solid line). As shown in \eqref{eq:final_Q}, the potential value is the limit of integration of the hot electron distribution function and therefor the chatge is highly dependent on this parameter. In fact, the increase in the potential barrier inhibits the  ejection of electrons and limits the charging process. This strangled regime appears for small targets as they concentrate the surfacic charge and enhance the background potential. This strangled regime also applies for low laser intensities because of a low value of $\phi_{min}$. As the ejection current strangling appears when $\phi_{bg} = \phi_{min}$, we use this relation to define the target size where the strangling effect is suppressed:
\begin{equation}
d_{min} =\frac{Q}{2\pi\epsilon_0 \phi_{min}}. \label{eq:d_c}
\end{equation}   
If the target is too small, we can also make a charging prediction, assuming an instantaneous struggling:
\begin{equation}
Q_{bg} = 2\pi d \epsilon_0 \phi_{min}.  \label{eq:Q_bg}
\end{equation}
In our experiments, the largest critical size is $d_{min} = 5$~mm, which is still smaller than the target size. While the strangled regime does not apply to our experiment, it was observed in Ref. \cite{Eder2009} and in Ref. \cite{Chen2012} where $d_{min}$ was larger than the target diameter of $1$~cm. Consequently, the authors measured a GEMP amplitude proportional to the target diameter, in agreement with Eq. (\ref{eq:Q_bg}).

\begin{figure}
\begin{centering}
\includegraphics[height=6.5cm]{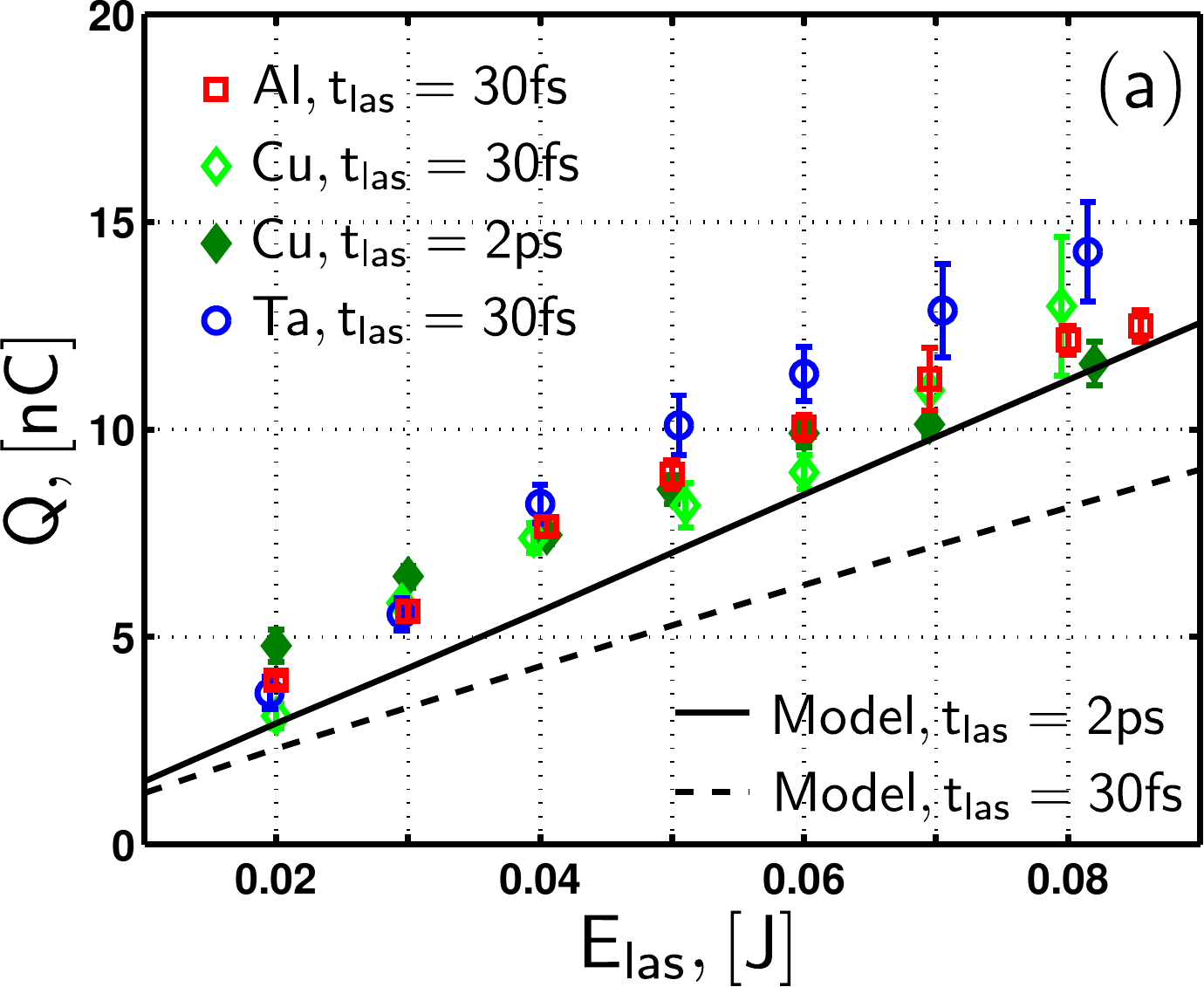}
\includegraphics[height=6.5cm]{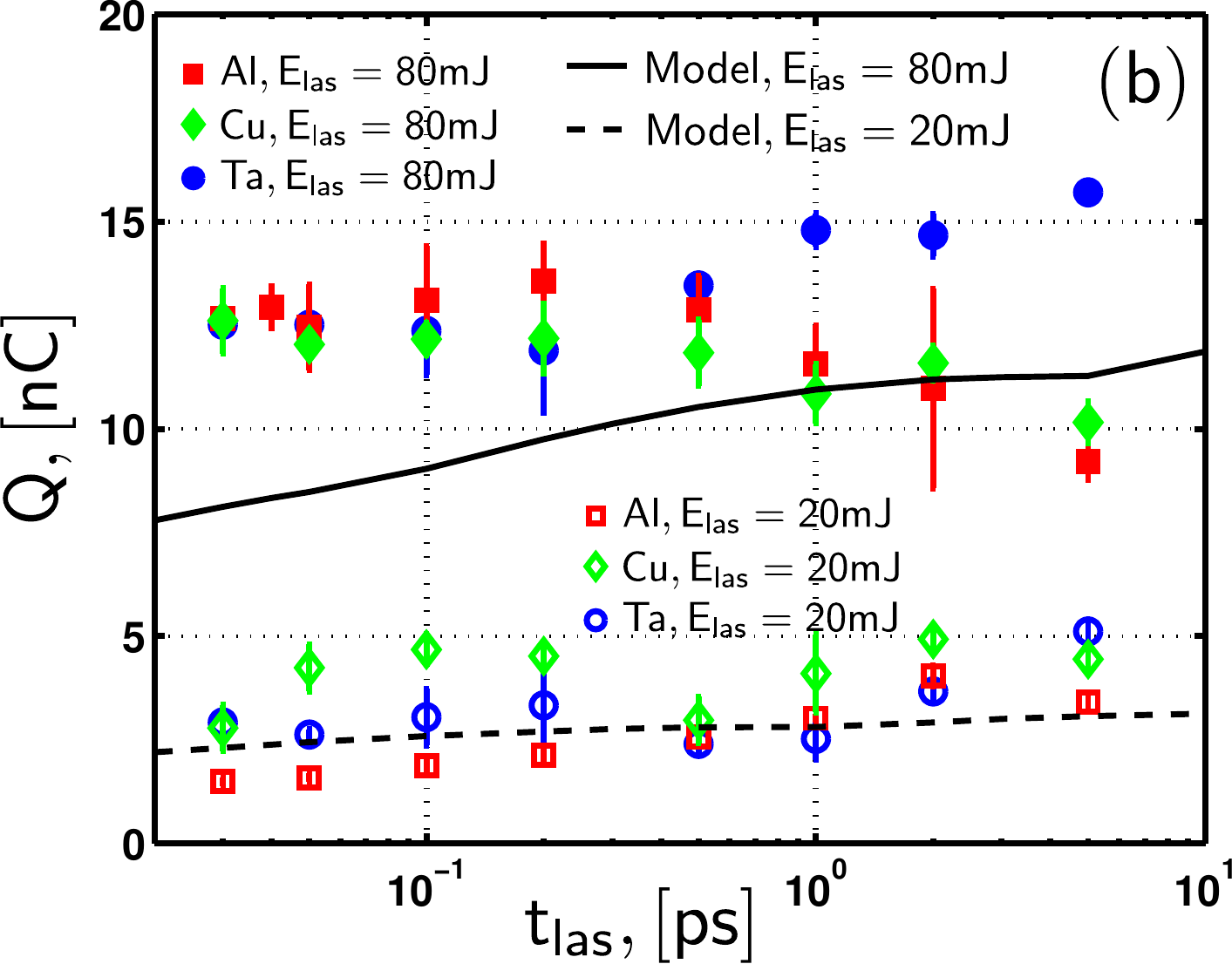}
\par\end{centering}
\caption{Target charge in nC measured experimentally (dots) and calculated with the model (line) as a function of laser energy  $E_{\rm las}$ (a) and pulse duration $t_{\rm las}$ (b),  for different target materials. The points are averaged over 5 shots. The error bars are the standard deviation on these points.\label{fig:exp_scans}}
\end{figure}
\section{Comparison between model and experiments}
A comparison between the measurements and the model predictions is shown in Fig. \ref{fig:exp_scans}a for scans of laser energy $E_{\rm las}$ and in Fig. \ref{fig:exp_scans}b for scans of pulse duration $t_{\rm las}$. The first important result is that the charge is independent of the target material. This feature is especially obvious at low laser energy and is implicit in the model, which does not include any target parameters (with the exception of the absorption coefficient).
The second result is the relative independence of the charge with respect to the laser pulse duration. A physical explanation of this independence can be seen undertood if we focus on the number of accelerated electrons given by (\ref{eq:N-accelerated}): for short pulses, the escaping electrons represent a large fraction of the small total number $N_e$ of hot electrons, while for longer pulses, they are a small fraction of the numerous hot electrons. Also, it should be noted that the charge scales linearly with $E_{\rm las}$ as in \cite{Brown2013}.
The model's predictions are in reasonable agreement with the experimental data for a broad range of laser parameters, considering the simplicity of the model. Nevertheless, there is some behaviour linked to the target material, for pulses longer than 0.5~ps, as illustrated in Fig. \ref{fig:exp_scans}b. As the material affects the absorption coefficient, this dependence may be accounted for by an appropriate variation of the laser energy absorption, taking into account a broad range of pulse durations.

\begin{figure}
\begin{centering}
\includegraphics[height=6.4cm]{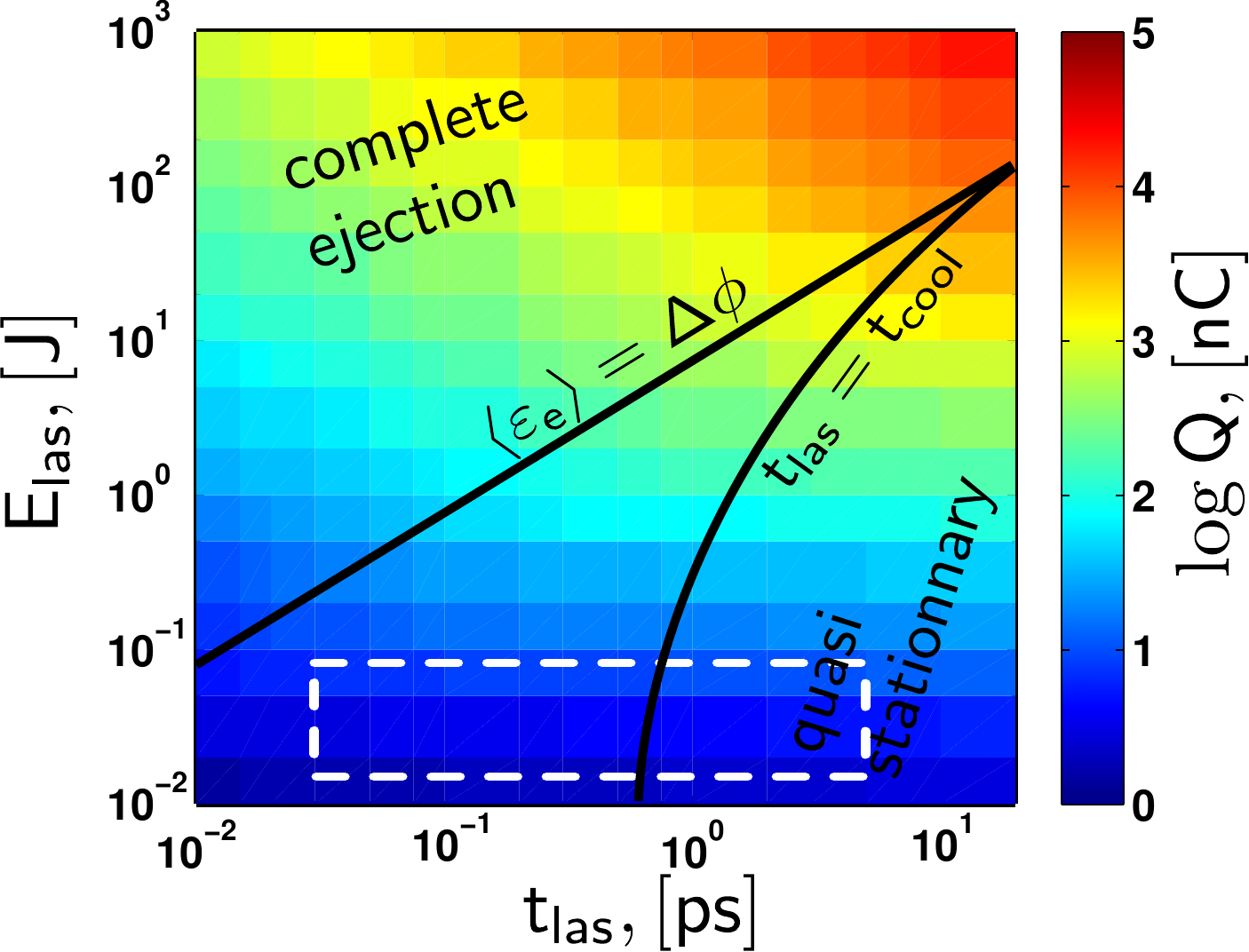}
\par\end{centering}
\caption{Expected target charge $Q_t$ in nC and in the logarithmic scale calculated from the model in function of the laser energy and the pulse duration for $r_{\rm las}=6$~\textmu\-m, $\eta_{\rm abs}=0.4$, and  $\lambda_{\rm las}=0.8$~\textmu\-m. The dashed line shows the domain explored in the present experiment.\label{fig:map_Q_Cu}}
\end{figure}

While the model is only corroborated experimentally for low laser energies, the model's assumptions are valid in a broader laser parameter space. Thus we extend the predictions of accumulated charge $Q$ as a function of laser energy and pulse duration in  Fig. \ref{fig:map_Q_Cu}, for a laser focal spot radius of $6$~\textmu\-m and an absorption coefficient of 40\%. These results should be carefully considered at high laser energy far from the validated range. This map highlights two different regimes of target charging. First, an almost complete hot electron ejection takes place if $\left<\varepsilon_e\right> \ge\Delta\phi_{th}$. In this case the target charge can be approximated by $Q\simeq eN_e$. Second, there is a quasi-stationary regime where the laser pulse duration is longer than the hot electron cooling time \cite{Dubois2014}. In this case, the current of ejected electrons is equal to $I_e = Q/t_{las}$.

\section{Conclusion}
In conclusion, by comparing our analytical model with our experimental results, we demonstrate that GEMP emission proceeds in two steps: the target is charged by the laser and is subsequently  discharged through the target holder. \\
In consequence, an ideally isolated target cannot generate GEMP in the GHz domain, even though it accumulates a charge. It may, however, produce TeraHertz waves due to the ejected electrons, x-rays due to the target plasma, or weak GEMP due to the polarisation current in the dielectric holder. Our second conclusion is that this simple fact allows us to determine two kind of laser-matter interaction experiments. If the charging time is shorter than the neutralisation time, there is a charge accumulation and then a GEMP emission. When these characteristic time are inversed, there is no charge accumulation but rather a constant current, which generates weak GEMP. Assuming a neutralisation time of $l/c$, a charging time proportional to $t_{las}$ and the holder size around $10$~cm, laser pulses shorter than $300$~ps produce GEMP with a strength proportional to the laser energy, while laser pulses longer than $300$~ps generate GEMP strongly inhibited by the neutralisation current. Our experiment produced GEMP efficiently because the ps and sub-ps laser pulses were much shorter than the characteristic discharge time (around 0.1 ns). Moreover, by properly shaping the target holder, it is possible to control the GEMP spectrum for diagnostics protection or for generation of strong quasi-static magnetic fields \cite{Fujioka2013,Santos2014}.

We acknowledge financial support from the French National Research Agency (ANR) in the frame work of "the investments for the future" Programme IdEx Bordeaux-LAPHIA (ANR-10-IDEX-03-02).
This work has been carried out within the framework of the EUROfusion Consortium and has received funding from the European Union's Horizon 2020 research and innovation programme under grant agreement number 633053. The views and opinions expressed herein do not necessarily reflect those of the European Commission.


\begin{thebibliography}{References}

\bibitem[1]{Pearlman1977} J. S. Pearlman \textit{et al.}, Appl. Phys. Lett. \textbf{31}, 414 (1977)
\bibitem[2]{Courtois2009} C. Courtois \textit{et al.}, Phys. Plasmas \textbf{16}, 013105 (2009) 
\bibitem[3]{Mead2004} M. J. Mead \textit{et al.}, Rev. Sci. Instrum. \textbf{75}, 4225 (2004)
\bibitem[4]{Brown2008} C.G. Brown Jr. \textit{et al.}, J. Phys. Conf. Ser. \textbf{112}, 032025 (2008)

\bibitem[5]{Fujioka2013} S. Fujioka \textit{et al.}, Sci. Reports \textbf{3}, 1170 (2013)
\bibitem[6]{Santos2014} J. J. Santos \textit{et al.}, \textit{Pulsed kilo-Tesla magnetic field generation by laser and applications}, submitted to Phys. Plasmas (2014)
\bibitem[7]{Dubois2014} J.-L. Dubois \textit{et al.}, Phys. Rev. E \textbf{89}, 013102 (2014)
\bibitem[8]{movie} Aditional Material: Lollipop.avi

\bibitem[9]{Felber2005} F. S. Felber, Appl. Phys. Lett. \textbf{86}, 231501 (2005)
\bibitem[10]{Sagisaka2008} A. Sagisaka \textit{et al.}, Appl. Phys. B \textbf{90}, 373 (2008)

\bibitem[11]{Jackson1975} J. D. Jackson, \textit{Classical Electrodynamics} (Wiley, New York, 1975)


\bibitem[12]{Bateman2012} J. E. Bateman \textit{et al.}, RAL Technical Reports, RAL-TR-2012-005 (2012)

\bibitem[13]{Crow1975} J. E. Crow, J. Plasma. Physics. \textbf{14}, 65-76 (1975)
\bibitem[14]{Mora2003} P. Mora, Phys. Rev. Lett. \textbf{90}, 185002 (2003)
\bibitem[15]{Schreiber2006} J. Schreiber, Phys. Rev. Lett. \textbf{97}, 045005 (2006)


\bibitem[16]{Gibbon2005} P. Gibbon,  \textit{Short pulse laser interaction with matter, an Introduction}, World Sci. Publ. London (2005) 
\bibitem[18]{Beg1997} F. N. Beg \textit{et al.}, Phys. Plasmas \textbf{4}, 447 (1997)
\bibitem[19]{Wilks1992} S. C. Wilks \textit{et al.}, Phys. Rev. Lett. \textbf{69}, 1383 (1992) 
\bibitem[20]{Fabbro1985} R. Fabbro \textit{et al.}, Phys. Fluids.  \textbf{28}, 1463 (1985) 

\bibitem[21]{Carron1976} N. J. Carron \textit{et al.}, IEEE Trans. Nucl. Sci. \textbf{23}, 1986 (1976) 

\bibitem[22]{Brown2013} C. G. Brown Jr. \textit{et al.}, J. Phys. Conf. Ser. \textbf{59}, 08012 (2013)


\bibitem[23]{Eder2009} D. C. Eder \textit{et al.}, Lawrence Livermore National Laboratory Report, LLNL-TR-411183 (2009) 
\bibitem[24]{Chen2012} Z. Y. Chen \textit{et al.}, Phys Plasmas \textbf{19}, 113116 (2012) 


\end{thebibliography}
\end{document}